# Title: How is model-related uncertainty quantified and reported in different disciplines?

**Authors:** Emily G. Simmonds[1,2]*, Kwaku Peprah Adjei[1,2], Christoffer Wold Andersen[3], Janne Cathrin Hetle Aspheim[1,2], Claudia Battistin[4,5], Nicola Bulso[4,5], Hannah Christensen[6], Benjamin Cretois[7], Ryan Cubero[4,5], Iván A. Davidovich[4,5], Lisa Dickel[8,2], Benjamin Dunn[1,4], Etienne Dunn-Sigouin[9,10], Karin Dyrstad[3], Sigurd Einum[2], Donata Giglio[11], Haakon Gjerløw[12], Amélie Godefroidt[3], Ricardo González-Gil[13], Soledad Gonzalo Cogno[4,5], Fabian Große[13,14], Paul Halloran[15], Mari F. Jensen[16,10], John James Kennedy[17], Peter Egge Langsæther[18], Jack H. Laverick[13], Debora Lederberger[19], Camille Li[9,10], Elizabeth Mandeville[20], Caitlin Mandeville[21,2], Espen Moe[3], Tobias Navarro Schröder[4], David Nunan[22], Jorge Sicacha Parada[1,2], Melanie Rae Simpson[23,24], Emma Sofie Skarstein[1], Clemens Spensberger[9,10], Richard Stevens[22], Aneesh Subramanian[11], Lea Svendsen[9], Ole Magnus Theisen[3], Connor Watret[13], Robert B. O'Hara[1,2]

**Affiliations:**
[1]Department of Mathematical Sciences, Norwegian University of Science and Technology.
[2]The Centre for Biodiversity Dynamics, Norwegian University of Science and Technology.
[3]Department of Sociology and Political Science, Norwegian University of Science and Technology.
[4]Kavli Institute for Systems Neuroscience, Norwegian University of Science and Technology.
[5]Centre for Neural Computation, Norwegian University of Science and Technology.
[6]Atmospheric, Oceanic and Planetary Physics, University of Oxford.
[7]Miljødate, Norwegian Institute for Nature Research.
[8]Department of Biology, Norwegian University of Science and Technology.
[9]Geophysical Institute, University of Bergen.
[10]Bjerknes Centre for Climate Research.
[11]University of Colorado Boulder.
[12]Peace Research Institute Oslo (PRIO).
[13]Department of Mathematics and Statistics, University of Strathclyde.
[14]Department of Ecology, Federal Institute of Hydrology, Germany.
[15]College of Life and Environmental Sciences, University of Exeter.
[16]Department of Earth Science, University of Bergen.
[17]Met Office, UK.
[18]Department of Political Science, University of Oslo.
[19]Schweizerisches Epilepsie Zentrum; Klinik Lengg, Zurich, Switzerland.
[20]College of Biological Science, University of Guelph.
[21]Department of Natural History, Norwegian University of Science and Technology.
[22]Nuffield Department of Primary Care Health Sciences, University of Oxford.
[23]Department of Public Health and Nursing, Norwegian University of Science and Technology.
[24]Clinical Research Unit Central Norway, St Olavs University Hospital.

*Corresponding author.

Email: emilygsimmonds@gmail.com
Postal address: Emily G Simmonds, Department of Biology, NTNU, NO-7491 Trondheim, Norway

## Abstract:

How do we know how much we know? Quantifying uncertainty associated with our modelling work is the only way we can answer how much we know about any phenomenon. With quantitative science now highly influential in the public sphere and the results from models translating into action, we must support our conclusions with sufficient rigour to produce useful, reproducible results. Incomplete consideration of model-based uncertainties can lead to false conclusions with real world impacts. Despite these potentially damaging consequences, uncertainty consideration is incomplete both within and across scientific fields. We take a unique interdisciplinary approach and conduct a systematic audit of model-related uncertainty quantification from seven scientific fields, spanning the biological, physical, and social sciences. Our results show no single field is achieving complete consideration of model uncertainties, but together we can fill the gaps. We propose opportunities to improve the quantification of uncertainty through use of a source framework for uncertainty consideration, model type specific guidelines, improved presentation, and shared best practice. We also identify shared outstanding challenges (uncertainty in input data, balancing trade-offs, error propagation, and defining how much uncertainty is required). Finally, we make nine concrete recommendations for current practice (following good practice guidelines and an uncertainty checklist, presenting uncertainty numerically, and propagating model-related uncertainty into conclusions), future research priorities (uncertainty in input data, quantifying uncertainty in complex models, and the importance of missing uncertainty in different contexts), and general research standards across the sciences (transparency about study limitations and dedicated uncertainty sections of manuscripts).

## Main Text:

Uncertainty is a well-acknowledged, fundamental part of the scientific process[1–5]. Uncertainty in scientific work can take myriad forms and is generated from a wide variety of sources. No universal taxonomy of uncertainty exists[6], despite many efforts to classify and categorise the diverse sources and forms of scientific uncertainty[3,7–11]. Generally, these taxonomies of uncertainty encompass three broad categories; uncertainty from natural randomness or variability in a system or process (aleatoric uncertainty), uncertainty in our knowledge of a system (including but not limited to; uncertainty in model structure, measurement and sampling errors, uncertainty in values of parameters), and uncertainty in our language, communication, and interpretation of processes. All of these sources are important contributors to scientific uncertainty, however, they cannot all be either quantified or reduced. In this paper, we focus on the second category, uncertainty in a system or process, refining further to concentrate on quantifiable uncertainty associated with the use of statistical or mathematical models (model-related uncertainty).

The importance of uncertainty associated with results of statistical and mathematical models is increasingly recognised because of prominent work such as climate change[3,12] and epidemiological models[13–16]. Nevertheless, quantification of model-related uncertainty and its reporting are not consistent or complete[2,15] within[3,5,15,17] or between scientific fields[1,18,19]. Despite similarities in descriptions of model-related uncertainty[3,8,20], a fully coherent picture has not emerged and different papers use different taxonomies of uncertainty and focus on different sources. There have been several calls for more consideration of uncertainty from specific fields or pairs of fields[1,3,5,19]. But these have yet to be answered comprehensively. This inconsistency can lead to confusion as to the true level of uncertainty in results and hinder interpretability across fields.

With quantitative science now highly influential in the public sphere[3] and the results from models translating into action, we must support our conclusions with sufficient rigour. Incomplete consideration of model uncertainties can lead to false conclusions with real world impacts and an erosion of public trust in science[15,17,21]. In 2019, Seibold et al[22] reported substantial declines in insect populations in Germany. This finding was widely publicised as an 'insect Armageddon'[23]. However, recent work by Daskalova et al[17] showed that a failure to account for uncertainty in model structure inflated confidence in the estimated declines. Only one of the five reported arthropod declines remained clear after uncertainty was corrected[17]. In 2020, epidemiological models were at the forefront of strategies related to COVID-19. An over-reliance on communicating point estimates/predictions masked the full range of possible outcomes and potentially contributed to slow or inappropriate government action[15,16,21].

All potential sources of uncertainty should be considered and accounted for when constructing, running, and interpreting statistical and mathematical models. The framework we use for our audit breaks model-related uncertainty into three primary sources: data (both observed and simulated), parameters, and model structure. The data element is further split into two sub-sources: the response, i.e. the focal variable trying to be explained, and the explanatory variables, i.e. any variables used to explain the response. This gives four sources in total to assess. An example of the framework as applied to a simple linear regression is given in the box 1. This 'source framework' is broad enough to be applicable to multiple scientific fields, while still capturing the main sources of model uncertainty.

**Box 1: Example of source framework**

Focal model: a simple linear regression of change in height of plants as a function of temperature.

Model equation:

$$\Delta H_i = \beta_0 + \beta_1 Temp + \varepsilon_i$$

$$\varepsilon_i = N(0, \sigma^2)$$

| Source | Element in the focal model | Example of potential uncertainty |
| --- | --- | --- |
| Response variable | Change in height ($\Delta H$) | Measurement/observation error |
| Explanatory variable | Temperature ($Temp$) | Measurement/observation error |
| Parameter estimates | Estimates of: Intercept ($\hat{\beta}_0$), slope of relationship ($\hat{\beta}_1$), and variance of the error ($\hat{\sigma}^2$) | Standard error/confidence interval |
| Model structure | The structure of the equation | Comparison of alternative formulations e.g. non-linear structure or additional explanatory variables |

Previous work has suggested that our current consideration of model uncertainty in the sciences is not sufficient[1–3,5,17], but the actual state of quantification and reporting in publications has not been assessed. To address this, we take a snapshot of the state of uncertainty reporting from papers published at the end of 2019 across seven scientific fields (final paper remaining in the analysis, N = 75 for Climate Science, 92 for Ecology, 57 for Evolution, 36 for Health Science, 89 for Neuroscience, 54 for Oceanography, and 93 for Political Science, Total = 496) to evaluate how they quantify and report model uncertainty in the key sources outlined above (detailed methodology will be provided in an online repository – attached here as additional file for reviewers).

*How well are we currently doing?*

The results of our snapshot assessment show that no field currently has a complete and consistent consideration of their model uncertainties (see Figure 1). However, across fields we get much closer to achieving this, offering opportunities for improvement; all four sources of uncertainty are quantified in at least 25% of instances within at least one field, with three sources having 50% or greater quantification in at least one field. Fields with low reporting of particular sources of uncertainty can learn from fields with high reporting of those sources. The one area where all fields fail to quantify uncertainty the majority of the time is that from explanatory variables. The fields that perform best here are Oceanography and Climate Science, each reporting uncertainty in about one quarter of papers assessed.

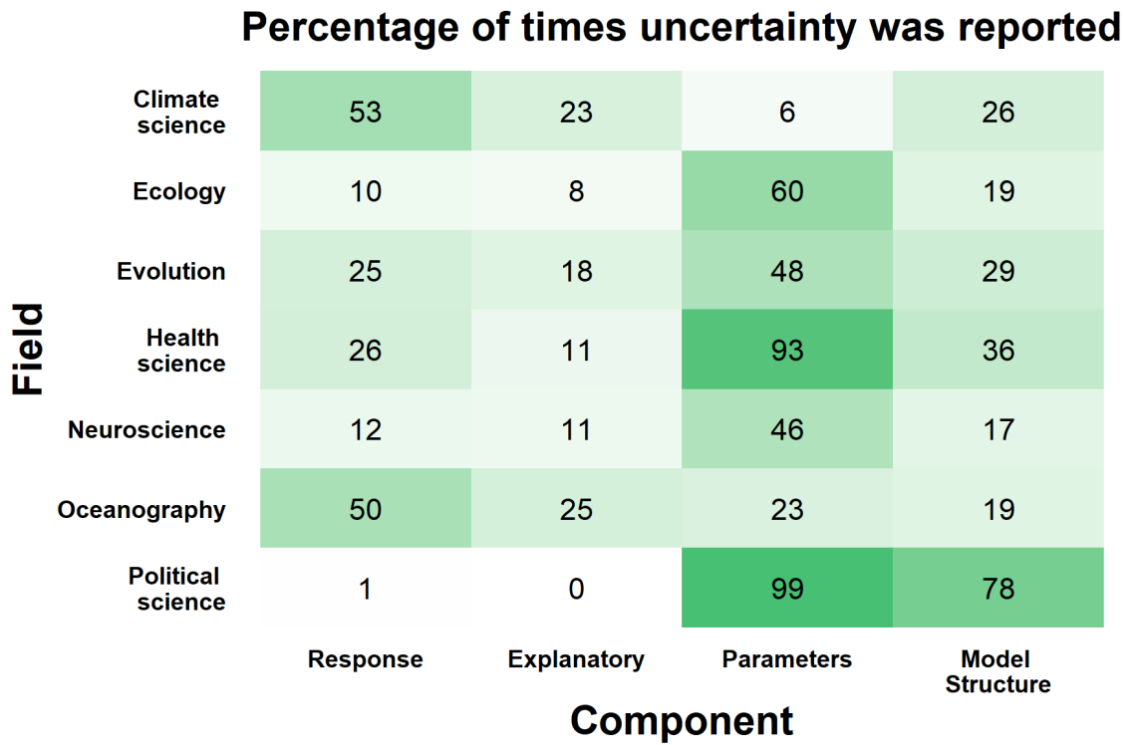

FIGURE 1: *Heat map of the percentage of papers that report uncertainty from each source split by field. Positive results include both papers that quantified and reported uncertainty arising from each source and those where the source introduced no uncertainty to the model. Only one focal model was considered per paper assessed. (N = 72 for Climate Science, 85 for Ecology, 52 for Evolution, 29 for Health Science, 87 for Neuroscience, 36 for Oceanography, and 76 for Political Science)*

We note that not all lack of quantification or reporting of uncertainty in a particular source represent an omission. There are some cases where quantifying uncertainty from a particular source is impossible, impractical, or unnecessary. One example of a nuanced requirement to explicitly quantify uncertainty is for response data in statistical models. Commonly applied statistical methods based on linear models, such as linear regression and ANOVA, do account for uncertainty in the response when estimating uncertainty in parameter estimates. However, they do not report it explicitly. Generally, this lack of reporting does not matter and would not influence results because it is the relationship between the explanatory variables and the response that is of interest scientifically. In other situations, it is necessary to explicitly quantify uncertainty in the response of statistical models. For example, survival of wild animals is typically derived from capture-recapture data, while its proper estimation would require explicit estimation of both the recapture process (observation) and a survival process. There are also cases where no applicable uncertainty is introduced from particular sources (see Figure 2). For example, in experimental studies, the explanatory variable is often a treatment group. These treatment groups are rarely a source of uncertainty in the modelling process, as group membership and treatment conditions are often known with certainty. Large numbers of experimental studies like this are present in Evolution and Health Sciences, and increasingly also in Political Science. A final example to note is when explanatory variables with noise or measurement error are actually the variable you want to represent such as in cases where explanatory variables are used for diagnosis or prognosis. In this case, it is the observed values of the explanatory variable which will be used for clinical use rather than the 'true' values and representing the uncertainty between observations and true values would be unnecessary.

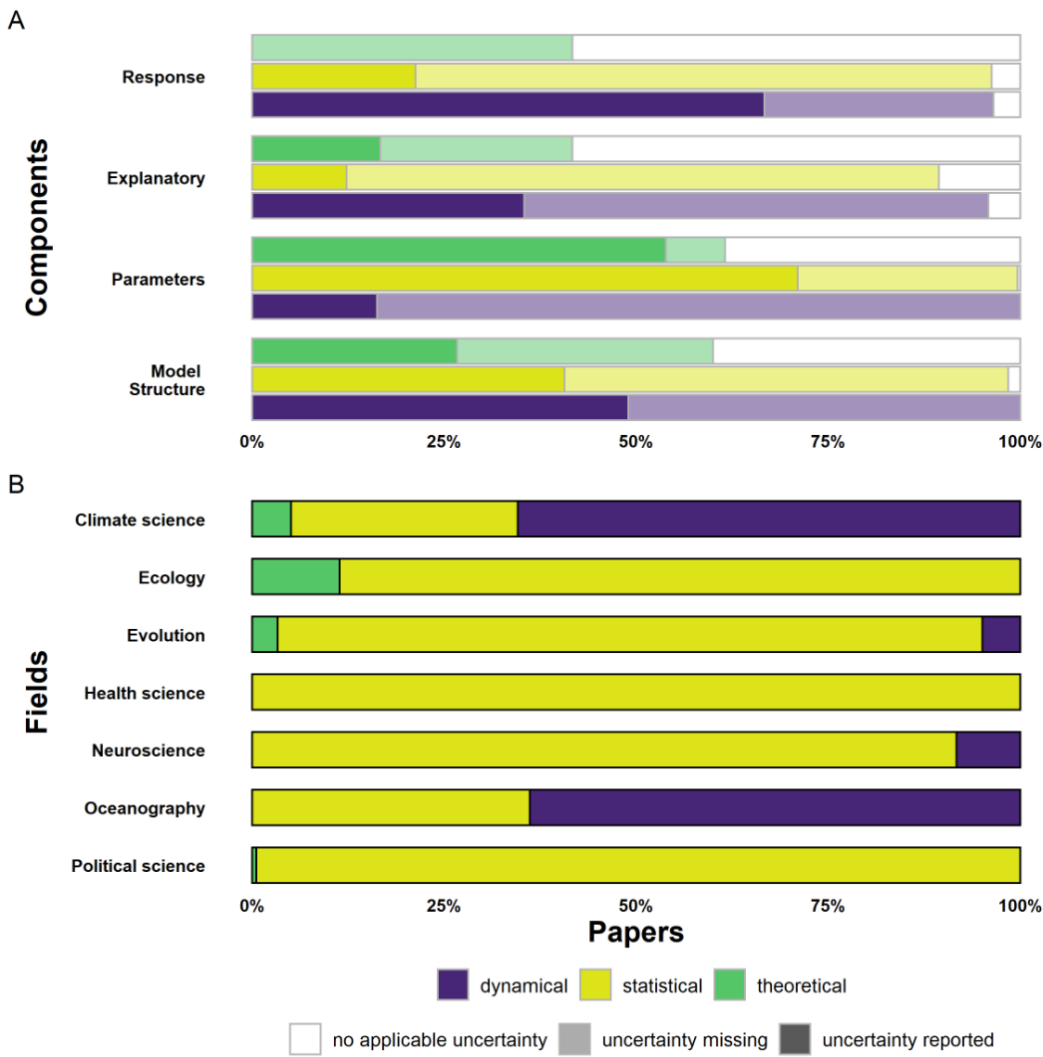

*FIGURE 2: A. Presentation of the percentage of models in which there was uncertainty reported, uncertainty missing or no applicable uncertainty for each source component (uncertainty was deemed not applicable if either the component was not relevant to the model or if there was no uncertainty in that component). (N = 57 for dynamical models (including mechanistic), 241 for statistical models, 12 for theoretical models or hybrid statistical/theoretical models). B. Percentage of model type assessed by field.*

Which sources of uncertainty were quantified and reported varied between fields but also by model type (see Figure 2). We classified all models in the audited papers into three broad model types: dynamical (a mathematical model based on fundamental understanding of natural processes such as physical or biochemical laws), statistical (a mathematical model that represents a data generation process, e.g. linear regression), and theoretical (a mathematical model designed to illustrate or test a theoretical idea, typically does not include observed data). Variation was found in how often uncertainty was quantified as well as whether any uncertainty was applicable for each source across the model types (Fig 2A). While most sources in our framework could contribute uncertainty to the majority of models, theoretical models (including hybrid statistical/theoretical models) were more likely to have source components with no applicable uncertainty. The differences in uncertainty relevance and quantification by model type align partially with the boundaries of the scientific fields considered (see Fig. 2B) as some fields focus more on dynamical and mechanistic models (Climate Science and Oceanography), while others rely more on statistical models (Health Science, Neuroscience, and Political Science), and some fields are more mixed (Ecology and Evolution).

We suggest that differences in uncertainty quantification are driven by differing perceived importance of the sources of uncertainty for each model type and for specific research questions, as well as by practical considerations. For example, parameter estimate uncertainty receives the most consistent acknowledgement for statistical models. This is likely because a fundamental aim of statistical analyses and tools is to estimate and draw inference from unknown parameters, and common standards exist for quantifying their uncertainty. In contrast, the aim of dynamical models is often to predict a response. In this case, uncertainty in the response was quantified most consistently, representing its greater focal importance for this model type in addition to the impracticality of quantifying uncertainty in the huge numbers of parameters and explanatory variables in complex geoscientific models. Additional gaps in uncertainty quantification are driven by lack of tools or guidelines for quantification associated with particular model types and by author omissions and trade-offs. We present below some opportunities to improve our practice based on our interdisciplinary insight into these challenges.

**Cross-disciplinary collaboration highlights opportunities for improvement**

Working with a large interdisciplinary team and informed by the results of our assessment of current practices, we identified several opportunities for improvement, summarised in Table 1.

*TABLE 1: Table of identified opportunities to improve uncertainty quantification and reporting, including details of the improvement.*

| *Identified opportunity* | *Detail* | *Exemplary fields* | *Fields that can benefit* |
|---|---|---|---|
| *Greater consistency* | *Use overarching source framework to identify potential routes for uncertainty to enter models, then follow model type specific guidelines of good practice for quantifying and considering these sources* | *For statistical models: Ecology, Evolution, Health Science, Neuroscience, Political Science*<br><br>*For dynamical and mechanistic models: Climate Science, Oceanography*<br><br>*For theoretical models: Ecology*<br><br>*For qualitative models: Climate Science, Political Science* | *Fields that use multiple model types: Climate Science, Ecology, Oceanography*<br><br>*Fields that are not yet as consistent in reporting: All* |
| *More complete uncertainty consideration* | *Share proposed good practice methods for quantifying uncertainty from different sources across model types and fields. See our guidelines and examples in Box 2.* | *Response: Climate Science, Oceanography*<br><br>*Explanatory: Evolution, Oceanography*<br><br>*Parameters: Health Science, Political Science*<br><br>*Model Structure: Political Science* | *All fields* |
| *Effective presentation* | *Recommended minimum numeric presentation to aid reproducibility and reusability of results and reduce ambiguity. Also recommend combining with visual presentation when feasible to aid interpretation.* | *Strong visual presenters: Climate Science, Ecology, Evolution, Neuroscience*<br><br>*Strong numeric presenters: Climate Science, Ecology, Oceanography, Political Science*<br><br>*Strong across all: Health Science* | *All fields* |

**Greater consistency through a common framework.** Achieving consistency in the quantification and reporting of model-related uncertainty across scientific fields is a challenging aim. Cross-discipline harmony has been hindered by both the lack of a standardised framework for considering model-related uncertainty and by field-specific vocabularies and different compositions of model types. Here we propose three complementary solutions which can help researchers address these challenges and produce easier cross-disciplinary comparisons. The first solution is a broad framework and common language through which to consider model-related uncertainty. We propose the use of our source framework as a tool to identify potential routes for uncertainty to enter the modelling process before referring to model type specific criteria for quantifying or addressing those uncertainties. The usability of this framework across different disciplines has already proven itself through our analysis.

The second solution we propose is to follow cross-disciplinary good practice guidelines, which we present in Box 2. Our audit notes that differences in model uncertainty quantification seem to be driven by model type rather than purely by scientific field (see Fig. 2B). As the sources of uncertainty in a particular model type are likely to be consistent across fields, we propose a logical split for guidelines of good practice by model type rather than scientific field. By focussing on model type and providing guidelines of good practice, as detailed in Box 2 for a statistical model, it is possible to achieve greater consistency and completeness in model-related uncertainty quantification across all scientific fields. Dialogue across fields is key to achieving greater consistency.

**More complete uncertainty consideration**. Across all the considered fields, we have documented the quantification of four sources of model uncertainty. However, no field or model type alone achieves this. Our proposed good practice guidelines (with specific examples in Box 2), informed by cross-field examples and practices, can support more complete uncertainty consideration for all models used across the sciences. For each potential source of uncertainty for each model type, we give indicate how uncertainty could be quantified, which in turn could be adapted to the model at hand.

The guidelines in Box 1 leverage good practice from fields with specialised modelling repertoires to create a comprehensive set of uncertainty practices that is relevant to all fields, particularly those using a diverse modelling repertoire. An example of good practice from specialised fields is the quantification of uncertainty from parameter estimation in Political and Health Science. Both fields have standard presentation styles for this type of uncertainty, which is most commonly associated with some form of regression analysis. In Political Science, standard practice is to report standard errors (or occasionally t-values) for regression coefficients. It is also common to visualise uncertainty by plotting coefficients or marginal effects with 95% confidence intervals. The numerical and or visual presentation of 95% confidence intervals is standard practice in Health Sciences. These accepted standards have led to an almost 100% success rate in reporting uncertainty from parameter estimates in the papers we assessed for Political and Health Sciences (see Figure 1). In contrast, in Ecology, despite also using predominantly statistical models, there is no such universal standard. Ecology research employs many different modelling tools and software platforms or packages, including a number of user-defined models. This lack of a consistent standard results in only a 60% reporting rate of parameter uncertainty in Ecology. Implementing clearer minimum expectations from statistical models like in Political or Health Science (as detailed in Box 2) could improve quantification of parameter uncertainty in Ecology and other fields. We have based our recommended methods on currently available tools and good practice from the seven considered fields, but we see these as working guidelines that should be updated as new tools become available.

One area that should be addressed in greater detail is how to deal with uncertainty in a particular source when quantification is impossible or unnecessary. This frequently occurs and can arise for many reasons including complexity of the model, logistical constraints, or methodological constraints. In these situations, applying our framework could be a useful method of systematically identifying what potential sources of uncertainty are missing from the current model, whether they could be incorporated or if they are quantified elsewhere. Researchers should subsequently consider how failure to quantify uncertainty is likely to affect the study conclusions.

Emphasis should be put on quality over quantity when applying these good practice guidelines. In theory, it could be possible to tick all boxes of addressing uncertainty from the four sources, without ever quantifying them correctly or thoroughly. Quality of the methods used to quantify uncertainty or how it was reported is not something we addressed in our assessment of papers; however, it is something that should be considered in future assessments and when developing good practices.

**Effective presentation of uncertainty**. It is not sufficient to only quantify model-based uncertainty; it is also essential to communicate it. Model-based uncertainty can be communicated in three primary ways: numerically, visually, or narratively. Across fields, we used different combinations of these communication types (see Fig. 3), ranging from 64% visual-only communication in Neuroscience to > 90% communication including numeric values for Health and Political Sciences to a balanced use of all communication types individually and in combination in Evolution and Oceanography.

While there exists a wide literature base discussing the most effective ways to communicate uncertainty, these papers often focus on a non-academic audience of policy makers or the public[24–28]. Their findings suggest that openly communicating bounded or quantified uncertainty can increase trust in results[15,21,26,29,30] and numeric and/or visual communication are more precise and effective than textual presentation for communicating the desired uncertainty level[2,25,31]. While we focus on communicating uncertainty to a primarily academic audience, several findings from this existing literature can also be useful for scientists. We propose a minimum presentation of quantified uncertainty as numeric values in scientific papers, either in the main text, in the supporting information; or in a supporting dataset published along with the paper. We recommend numeric presentation for two reasons, first, to reduce ambiguity that could come from textual or visual presentation[2,32,33] and second, to aid in reusability of results. Numeric presentation of uncertainty is essential for the reuse of results in systematic reviews or meta-analyses, or for reproducing the results in future and therefore is essential for the progress of research. We also recommend including visual presentation when feasible to aid interpretation[2,25,34]. There are many ways in which uncertainty can be presented visually, increasing the potential for an effective method to be found[2,33]. Presenting model results visually can aid with understanding of complex relationships but are not free from bias or misinterpretation, which is why we recommend a combination of numeric and visual communication[33,35]. We recommend as a best practice that the code used to produce uncertainty presentations is shared to enhance the replicability and transparency of uncertainty quantification.

To effectively implement our communication recommendations, we advise developing a set of standard uncertainty analyses and tools to implement them (either within existing software/packages or as post-processing steps) so that every modeller can generate uncertainty metrics for their work. This would allow easier production of visual or summary tabular representations of model-based uncertainties, which can be included in the main manuscript text. This should then be coupled with larger tables of numeric values in supporting information including full uncertainty bounds for each quantifiable source from the source framework. Again, we can take inspiration on good practice for uncertainty communication from some of the fields included in our audit. For example, Health Science presents uncertainty using numeric, visual, and text methods 40% of the time and includes some numeric representation > 95% of the time. An example of good practice we encountered was Heisser et al. [36] who coupled visual presentation with numeric bounds in their Figure 3. In contrast, other fields such as Climate Science, Evolution, and Neuroscience report uncertainty numerically less than 50% of the time. These fields can learn from standard practice and examples from Health Science to improve their own uncertainty communication. The transferability of good practice will depend somewhat on the model used, for example, visual presentation of parameter uncertainty for a model with greater than 100 parameters will not be practical. However, inspiration can still be taken to improve communication of uncertainty when presentation is practical.

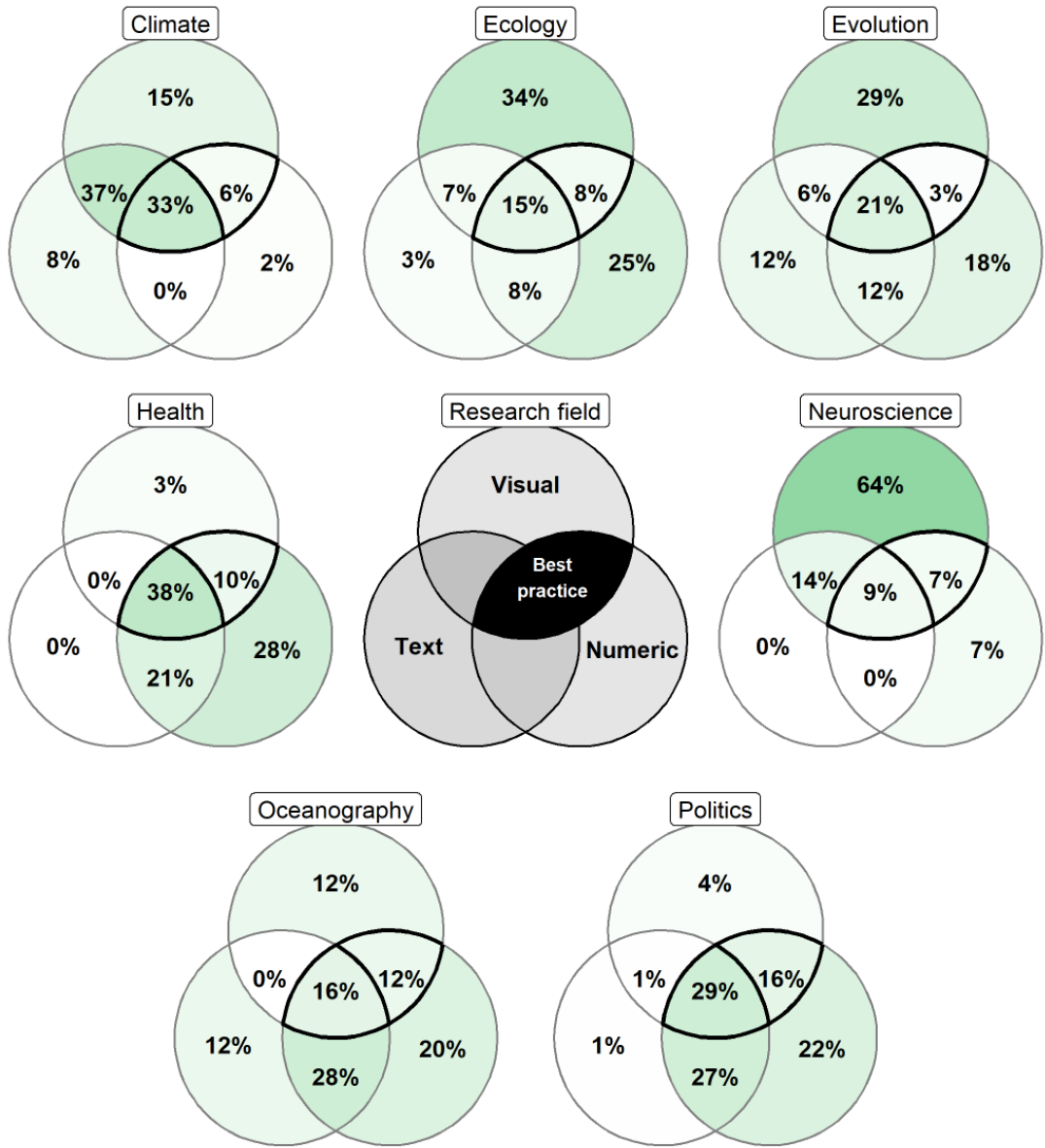

FIGURE 3: Venn diagrams of presentation methods for uncertainty (legend in the centre) by field. Black edged segments are our good practice recommendation of visual + numeric or visual + numeric + text.

*Box 2: GOOD PRACTICE GUIDELINES FOR ALL MODELS TYPES including CASE STUDY implementation for each uncertainty source as identified during our assessment for statistical models.*

**Good practice common to all model types:**

- Compare, contrast, or represent (through averaging of parameter estimates) the results of alternative **model structures**
- Discuss any uncertainty sources that were not quantified explicitly in the main text or discussion section and, explain why and detail how this could impact the results and conclusions of the analysis. Include a dedicated 'uncertainty' section of the paper (possibly in supporting information)
- Publish code and/or data/model output used for the analyses to ensure work is **reproducible and reusable**

**Statistical model good practice**:
- Quantify any error in the **response** data. This can be in the form of explicit modelling of measurement/observation error and subsequent correction, correction for non-independence, or an estimation of the error/bias that is propagated into the focal model
- Quantify any error in the **explanatory variable** data. This can be in the form of explicit modelling of measurement/observation error and subsequent correction, correction for non-independence or confounding variables, an ensemble approach to represent multiple data sources, or an estimation of the error/bias that is propagated into the focal model. If explanatory variable data comes from another model output (as is the case for projections of future climate) the full uncertainty associated with this output should be propagated into the focal model
- Present error estimates or an interval representing the plausible **parameter** space for all unknown parameters. This could be as a confidence interval, credible interval, bootstrap interval, or standard errors

**Dynamical model good practice:**

- Quantify any uncertainty entering the model from the response, if necessary (i.e. when the response is not a predicted outcome of the model and the aim of the model is quantitative understanding). This can be in the form of reporting model parameters (e.g. probability of detection) or statistics such as repeatability
- Quantify any relevant uncertainty in the **explanatory variable** data. This can be realized by running model ensembles with perturbed fields of the explanatory variables (i.e model forcing or initial conditions in most cases) with the strongest influence on the studied response
- While assessing uncertainty in all parameters in a dynamical model can be unfeasible (especially for data intensive modelling such as climate science or oceanography), parameter uncertainty can be quantified similarly to that of explanatory variables by running model ensembles covering a range of possible values for the parameters in those equations with the strongest influence on the studied response
- In addition, for both uncertainty in explanatory variables and parameters, simplified versions of the models could be used for a more comprehensive uncertainty analysis. Although the uncertainty quantified via such an approach would not be identical to that of the original more complex model, it would provide a dependable estimate

**Theoretical model good practice:**

- Check if it is necessary to quantify uncertainty in the **response**. Quantify any uncertainty in the response data. Typically a response in a theoretical model will be the outcome of the model rather than an input and is therefore predicted. The response int this case is not strictly a source of uncertainty but it does accumulate uncertainty from all other sources. Therefore to correctly represent uncertainty in the response, it is necessary to present the response accounting for the uncertainty introduced from all other sources. This can be in the form of presenting intervals around predicted response values, presenting a distribution of response values, or a range or other summary statistics that include variability of the results. May not have uncertainty if it is a deterministic model.

- Quantify any relevant uncertainty in the **explanatory variable** data. Explanatory variable data can either come from observations, experiments, or be simulated as part of the model. Each form of explanatory variable is a source of uncertainty in a different way. This uncertainty can be quantified using measurement/observation error modelling, choosing a range of values/sampling values from a distribution during a simulation or bootstrapping or sensitivity analyses to assess the impact of changes in explanatory variable values.
- Parameter values in theoretical models are often chosen a priori or optimised using various algorithms. Often parameters are chosen specifically based on previous scientific findings, from observed data, from knowledge of physical or chemical processes, or to test a specific theory. Uncertainty in these parameters should be quantified by choosing a range of values/sampling values from a distribution during a simulation or bootstrapping or sensitivity analyses to assess the impact of changes in explanatory variable values. In some cases, there will be no uncertainty added from the unknown parameters because the question being asked is dependent on specific parameter values, for example, does an increase of 1ºC in mean sea surface temperatures cause greater carbon drawdown into food webs? In this case, the temperature parameters would need to be fixed and therefore would not be a source of uncertainty.
- Compare, contrast, or represent (through averaging of parameter estimates) the results of alternative **model structures**. This is not always relevant in theoretical models when they test if a specific model structure can produce the outcome expected. This is because the aim of many theoretical models is to test the model structure specifically. However, in other cases, where the motivation is simply to find a model that can represent reality, infinite different options could be available and some consideration of this breadth should be included.

Cross-disciplinary example implementations for **statistical models**:

| Field | Source of uncertainty and quantification metric | Citation | Details |
|---|---|---|---|
| 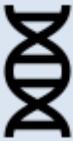 | 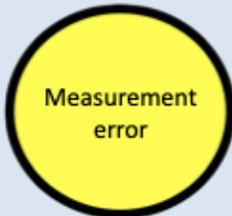 Measurement error | Conith *et al.* *Evolution.* 73, 2072–2084 (2019) | Response: Conith et al.[37] conducted a comparative analysis of the evolutionary history of snout length and depth for cichlid fish. They corrected their response variables of snout length and depth using a theoretically expected variance covariance matrix of correlation among traits which controls for non-independence from shared evolutionary history and taking account of the body depth of each specimen. This produced a 'corrected' response variable that was independent of size and history, thus resolving uncertainty in what morphological or evolutionary processes this variable represents.<br><br>**example of correction for non-independence** |
| 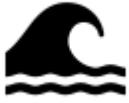 | 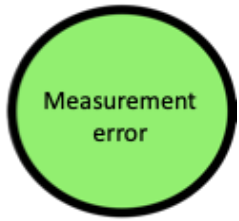 | Saderne *et al.* *J.Geophys. Res. Ocean.* 124, 7513–7528 (2019) | Explanatory: Saderne et al.[38] used a mechanistic model to explore differences in the $CO_2$ system across three ecosystems (coral reefs, mangroves, and seagrass meadows). One explanatory variable was pH on the total scale. This variable was corrected using extra data collected using a different method and the corrected variable used in further analyses. Other input data elements (which were used as parameters in the final model) had their errors propagated using the R package *Seacarb*[39]. This allowed for instrument error to be accounted for throughout the analysis. These techniques for considering and accounting for error in explanatory variables can and should be applied across statistical models as well.<br><br>**example of correction for measurement/ observation error** |
| 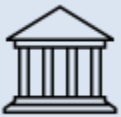 | 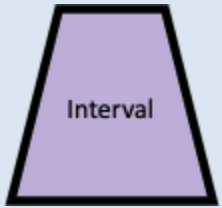 | O'Grady. *Br. J. Polit. Sci.* 49, 1381–1406 (2019) | Parameters: O'Grady[40] modelled preferences for increases in federal social spending with explanatory variables of household income and subjective assessment of unemployment risk. The model used was a multivariate regression. Uncertainty in the parameter estimates from this regression were calculated using standard errors and presented as clustered standard errors numerically in their Table 1 and then as 95% confidence intervals around the coefficient estimate visually in their Figure 5.<br><br>**example of error estimates using standard errors** |
| 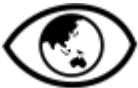 | 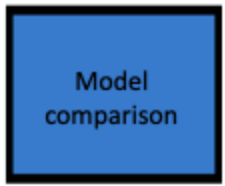 | Fan *et al. Clim. Dyn.* 53, 6919–6932 (2019) | Model structure: Fan et al.[41] compared two different parameterisations of a dynamical model for solar energy distribution and surface hydrology over the Tibetan Plateau. The dynamical model was the Community Climate System Model (CCSM4). One parameterisation used 3D radiative transfer and the other used a plane-parallel radiative transfer. The results of both parameterisations were compared visually in figures, as text, and numerically. Consideration of different processes and how they could impact results should also be considered in statistical models.<br><br>**example of comparing results using two model structures** |

## Outstanding challenges

In addition to the opportunities for improvement identified in the section above, we also note shared outstanding challenges to effective and comprehensive quantification of model-based uncertainty that span the included fields and model types. These challenges are yet to have satisfactory solutions for any field. We discuss each challenge in detail below and propose some next steps for the research community to create a path to overcome them.

**Input data uncertainty.** Across all fields assessed here, the source of uncertainty reported least often is for the data sources. Uncertainty from explanatory variables was quantified and reported < 37% of the time and uncertainty from the response was reported < 54% of the time across fields (see Figure 1). The slightly higher reporting rate of uncertainty in the response is driven by greater reporting for dynamical models (see Figure 2). This is because responses in these models are rarely in the form of input data, instead being predictions or validation data. It is the quantification and reporting of uncertainty arising from input data (observed data used as inputs to a model) that we identified as a particular challenge, but consequences of ignoring it can be severe[42]. Uncertainty can enter the modelling process from input data (both the response and explanatory variables) through random noise, unknown measurement or observation errors, or missed processes that could lead to imprecise predictions. We identified three key challenges that are preventing quantification of input data uncertainty. The first is logistics as it is not always possible to obtain sufficient measurements to be able to quantify uncertainty in particular variables. This is more common when data are very costly or logistically challenging to collect. The second and third barriers are related and comprise both a lack of knowledge of methods to quantify input data uncertainty and a lack of knowledge of the impact of failing to quantify input data uncertainty in different situations.

Tools do exist to quantify or remove most of these sources of uncertainty. There is a long history of using various error-in-variables models to account for uncertainty in the measurement of explanatory variables[43–46]. These models are supported by detailed theory in relation to linear models[46,47] and some non-linear models[43,48] and there are a myriad of options of models to quantify this type of uncertainty or bias. Models also exist to quantify observation error in response and explanatory variables, through the mapping of observed data to an unobserved or latent state (state-space models), which are widely used in demographic ecology[49]. Despite the availability of these tools, our results show that generally across all included fields, it is not standard practice to employ them. Wide scale implementation of such methods has been hindered by lack of knowledge of their practical applications, insufficient availability of data, no rules of thumb for when errors will influence results in many cases, and not including such considerations in standard statistical teaching. For response variables specifically, as mentioned in the introduction, many standard statistical models already adequately account for this uncertainty provided uncertainty in the response follows a normal distribution. This results in little reporting of uncertainty in the response for statistical models, which largely is not an omission. We would go so far as to not recommend any extra consideration for uncertainty in the response in these cases since it is adequately accounted for by standard practice. However, it is still important to note that it is not always the case that these assumptions are met. In some situations further consideration and explicit quantification may be required and yet omitted as it is not standard practice to consider uncertainty in the response when checking for deviations from the assumptions of the model.

We suggest a single solution to all three barriers to our consideration of uncertainty in input data. We call for further theoretical or simulation work exploring the impact of unquantified uncertainty in observed data in different contexts, especially when repeat data collection is challenging/impossible. This work should be coupled with better communication of the methods and easy platforms or packages to implement them. This final step would open accessibility to a wider range of scientists even without a formal statistical background.

**Model complexity/logistics (including machine learning and artificial intelligence).** Increased computing power, improvements in data collection technologies, and developments of machine learning and artificial intelligence (AI) have allowed us to develop more complex statistical models of hard to study systems and automatic algorithms to fit them. However, with these increases in complexity come trade-offs in terms of quantification of model-related uncertainty and in interpretability, with many of these complex models often being treated as a 'black box'. In parallel, complex models are frequently required in Climate Science and Oceanography to model the highly complex Earth system. Such models are often the only insight we can have into the behaviour of physical systems on earth and are the best we can achieve with current tools. Using these complex models is therefore a crucial step in the progress of science.

However, it can be impossible, due to practical limitations, to quantify uncertainty in all parameters or input data sources for highly complex models[50], for instance, the computing resources required to quantify sensitivities for each parameter in a highly complex model could be beyond what is currently achievable. In some cases, attempting to quantify uncertainty in all parameters can actually reduce the accuracy of the model, having the opposite effect to the intention[50]. A trade-off can therefore be made in deciding how to play off model resolution (and increased complexity) against the assessment of uncertainty. In many cases, being more thorough about uncertainty can mean doing worse in terms of accuracy – for example, increasing model resolution may resolve a new process (e.g. eddies in the ocean) which means the result jumps out of the region where an uncertainty analysis at lower resolution would have bounded the problem, but now falls close to where the real truth lies. Therefore, caution should be exercised when trying to address all sources of uncertainty in complex models and ensure we have the correct tools to achieve this successfully. The trade-off exists between what we can now model and what we can interpret and quantify uncertainty for.

We suggest two solutions to the trade-off between model complexity and uncertainty quantification. The first is to include a specific uncertainty section of all manuscripts potentially as a designated supporting information section. This uncertainty section would contain discussion of the limits and assumptions of the models in terms of uncertainty. A dedicated section would also give space to discuss the potential consequences of any unquantified uncertainty, including reductions in accuracy or giving indications of which elements could be expected to change. The second is to call for further research to improve methods for quantifying uncertainty in complex models and model fitting algorithms, ensuring uncertainty quantification keeps pace with model development.

**Propagation/within paper consistency.** During our audit, we noted two issues with propagation of model uncertainty. The first was within the models in a paper, where we observed that in some papers with multiple analyses, uncertainty consideration was not consistent across all models. This pattern was especially prominent for papers including multiple analysis types within a single study, a more common occurrence in fields such as Ecology, Evolution, and Neuroscience. The second was in propagation of uncertainty in results into the final discussion and conclusions. Rarely did we find that quantified uncertainty was propagated through into the discussion and conclusion of manuscripts. This pattern was universal across all fields. Even when uncertainty was reported earlier in the manuscript, conclusions still were largely based on point estimates or mean patterns. Uncertainty that was mentioned in the discussion sometimes focused on missing processes or caveats to the conclusions rather than quantified uncertainty. Both of these propagation issues can hinder the interpretation of the uncertainty associated with results.

We propose using our good practice guidelines for different model types to ensure a consistency to uncertainty consideration across all models in papers. Using a source based framework for these guidelines helps to identify where uncertainty enters the modelling process and therefore improves propagation of uncertainty. We also encourage journals to set an expectation for paper conclusions and discussion to include reference to model-based uncertainty. Currently, there can be trepidation among authors about diluting their conclusions by incorporating uncertainty, especially due to the highly competitive publishing and funding environment and the potential of public influence. Improving acceptance of transparent uncertainty in scientific conclusions would be a way forward to propagation into final results.

**Interdisciplinary working.** It is increasingly recognized that an interdisciplinary approach is required to address many of the key questions facing society, including climate change, the biodiversity crisis, and global pandemics. Much of this work revolves around statistical or mathematical modelling, often integrating approaches from multiple disciplines. Quantification of uncertainty is essential for the effective application of this work to societal decision-making, but the lack of a common framework for understanding uncertainty across fields makes it difficult to assess uncertainty in complex multidisciplinary systems [51]. We propose that our push for greater consistency in the quantification of uncertainty across fields will facilitate better reporting of uncertainty in interdisciplinary work, which we expect will aid interpretation and application by multiple audiences

**How much uncertainty consideration is enough?** It is not possible to account for all possible uncertainty in our studies[5], there will always be unknown unknowns that remain and known sources that for practical and well considered reasons cannot be addressed in a given study. However, we do need to ensure that we make best efforts to represent as much of the uncertainty related to our results as we can and in the correct way. How we can achieve accurate uncertainty representation without over complicating or diluting results, remains an open question. In relation to the framework we present, we need to ask: are there times when explicit quantification of all sources is actually not enough? Model-based uncertainty is not the only uncertainty associated with scientific work.

How much is enough will be case-specific and nuanced. It does not only depend on the type of model used but also the aims and focus of the work, for example a regression model of heart rate (response) as a function of age (explanatory) where the authors do not care about the distinction between 'true' and 'observed' heart rate would not require explicit quantification of uncertainty in the response. In this example, the main focus of the paper is on the relationship between age and heart rate (either observed or true) there is a strong need to be very sure of relationship and the uncertainty in that parameter and this need is satisfied because uncertainty from the response is accounted for in the estimation of the parameter and its uncertainty. In contrast, there could be a case where authors want to make conclusions specifically for the 'true' value not the 'observation' and make predictions for it; for example, if managing a population of endangered animals. Here it is essential to explicitly know how the observed counts map to true population values, therefore requiring explicit quantification of uncertainty in the response. It can be exceptionally challenging to tease apart these nuances from published papers because it is not always explicit how the authors view each model component or how generalisable they intend their work to be. Furthermore, the limits and assumptions of models used are frequently not discussed in sufficient detail.

We propose a solution to illuminate the aims and scope of different model-based analyses and encourage authors to discuss all uncertainty in their work, even those elements that are not quantified. This would be achieved by having a specific uncertainty section of all manuscripts, as also suggested to aid uncertainty reporting for complex models. This section could include specific author statements on which sources of uncertainty have been quantified and why, which elements are missing and what their impact might be, as well as the intended scope of the work and any limitations. This would help readers to appreciate the full uncertainty associated with the work and aid in correct reuse, replication, or citing of any results.

Some fears exist in the academic community about being explicit about uncertainty in our work, assuming that public or policy audiences might lose trust in results. Indeed, in some situations communicating uncertainty can influence public perceptions negatively due to ambiguity aversion, such as with vaccine effectiveness[52]. However, there have also been several findings indicating that public trust in scientific results is not reduced by communicating uncertainty[2]. Some have even shown the opposite, demonstrating that a lack of transparency around uncertainty can erode public trust[15,17,21,53]. Therefore, we should not shy away from reporting uncertainties associated with our work but instead ensure we communicate them as fully and transparently as possible and in an easy to interpret manner.

As mentioned in the introduction, uncertainty can enter the scientific process from a myriad of sources, with model-associated uncertainty being just one. As a result, it would be possible to score perfectly using our proposed source framework, quantifying all sources of model-based uncertainty in some way, but still have results that are subject to large unquantified uncertainties. Furthermore, in our analyses we made no judgement about whether a certain method of quantifying uncertainty was the most appropriate or what impact the quantification had on results; for example, whether results actually became less accurate in the pursuit of better quantified precision. Quality of uncertainty quantification will also be a crucial element in determining whether enough has been done in any given analysis. Simply ticking all boxes is not sufficient, good practice for each must also be followed (see Box 2)[16].

## Recommended ways forward

Our analysis reveals a lack of consistency in uncertainty quantification within and between fields. The fact that some fields do successfully account for uncertainties for certain model types and sources while others do not indicate that it is disciplinary protocols or customs that have led us to this state. Our analysis also highlights the potential for improvement. To reveal these previously unnoticed patterns, we had to translate the discipline-specific terminology surrounding model development and uncertainty quantification into a common language.

We make nine concrete recommendations for current practice, future work, and general research recommendations. The first two categories are aimed at the modelling community and the third is aimed more broadly, including at scientific publications and funders.

Recommendations for standard practice in quantifying model uncertainty:

1. Use the source framework as a structured tool for considering model uncertainty. Where uncertainties from the sources can and should be quantified, do so. Where it is not feasible or practical to quantify a particular source of uncertainty, instead include a theoretical discussion and acknowledgement of the missing uncertainty, why it is missing, and consideration of what impact it may have on the results reported.
2. Follow our proposed inter-disciplinary good practice guidelines for uncertainty quantification (see Box 2).
3. Present model uncertainty as clearly as possible using at minimum some numeric presentation to aid reuse and reduce ambiguity. Should be combined with visual presentation when feasible to aid interpretability.
4. Propagate model uncertainties into the conclusions drawn from the work.

Recommended future research priorities:

5. Develop tools and guidance on how to identify when uncertainty from input data is important.
6. Couple modelling advances such as increased complexity or AI and machine learning with further theoretical work on how to quantify and propagate the uncertainties associated with such methods.

7. Conduct further research into the influence and importance of the different sources of uncertainty for final results and conclusions across multiple modelling types and contexts.

General recommendations:

8. Where uncertainty cannot yet be quantified and its impact is not known, be transparent about these limitations, especially when drawing conclusions. Be accepting of conclusions that include explicit recognition of model uncertainty
9. Make it standard to have transparent and easy access to quantified model uncertainties in all manuscripts, e.g. through standard dedicated supporting information sections

Recommendations 1, 2, 3, 4, and 8 can be implemented immediately but recommendations 5, 6, 7 and 9 require long-term planning.

## Conflict of interest:

Authors declare that they have no competing interests.

## Acknowledgements:

We would like to acknowledge and thank those collaborators who left the team due to extenuating circumstances.

## Author contributions:

Conceptualization: EGS, RBOH

Methodology: EGS, RBOH, KPA, BC, RGG, JHL, CM, JSP, ESS

Audit of papers: All authors

Visualization: EGS, JHL

Project administration: EGS (whole project), CB (Neuroscience), EDS (Climate Science), FG (Oceanography)

Writing – original draft: All authors

Writing – review & editing: All authors

## Supplementary materials:

**This document contains**

**Section**

S1 The framework
S2 Development of Audit Strategy
S3 The Systematic Audit
S4 Detailed reviewer instructions

# S1 The framework

The framework used for this review is a source-based framework and was developed specifically for this study. It breaks model-related uncertainty into three primary sources: data (both observed and simulated), parameters, and model structure. The data element is further split into two sub-sources: the response, i.e. the focal variable trying to be explained, and the explanatory variables, i.e. any variables used to explain the response. This gives four sources in total to assess.

| SOURCE COMPONENT | DEFINITION | HOW IS IT A SOURCE OF UNCERTAINTY? |
|---|---|---|
| Response | The variable(s) of interest, the quantity(ies) we want to explain or predict. Could be something measured or something latent (unobserved) or simulated. | If the response is observed, it can have unknown error and biases. If the response is latent and therefore never observed, then it needs to be estimated. Responses can also be output from another model e.g. interpolated temperature. Predicted responses will have all uncertainty from predictive model. |
| Explanatory variables | Any variable that explains or predicts the response. Could be something measured or something latent (unobserved) or simulated or theoretical. | If the explanatory variables are observed, they can have unknown error and biases. If the covariates are latent and therefore never observed, then they need to be estimated. Covariates could also be output from another model e.g. interpolated temperature. Theoretical covariates could have many plausible values (e.g. initial conditions in a simulation). |
| Parameter estimates | Values given to unknown parameters in the model either through estimation, optimisation, or chosen. | Can arise from the statistical estimation of unknown parameter values, which gives the range of plausible parameter values generated from estimation. Can also arise from optimisation or selection of parameter values for a theoretical model. |
| Model structure | Uncertainty in the process being investigated – the structure of the equations that link the response, explanatory variables and parameters. Can be called 'model uncertainty' in some disciplines. | Different model structures contain different assumptions, parameters, and covariates. They therefore produce different results for the same response. |

Models typically contain all of the components of the framework. Nevertheless, one or several may be missing or minimally important in specific cases. Together, the components can be considered a modelling unit. As model complexity increases, the number of units also increases, leading to more layers of uncertainty (e.g. in a predictive model). However, we do not assess these layers specifically here.

*Example of the framework in practice:*

Focal model: a simple linear regression of change in height of plants as a function of temperature.

Model equation:

$$\Delta H_i = \beta_0 + \beta_1 \, Temp_i + \varepsilon_i$$

$$\varepsilon_i \sim N(0, \sigma^2)$$

| Source | Element in the focal model | Example of potential uncertainty | Example of quantification |
|---|---|---|---|
| 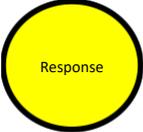 Response | Change in height ($\Delta H_i$), this element covers uncertainty in $\Delta H_i$ not $E(\Delta H)$ | Unknown observation error | Explicit model of the observation process e.g. $$y_i = \beta_0 + \beta_1 \, Temp_i + \varepsilon_i$$ $$\varepsilon_i \sim N(0, \sigma^2)$$ $$\Delta H_i \sim Pois(y_i)$$ Where $y_i$ is the true change in height and $\Delta H_i$ is the observed change |
| 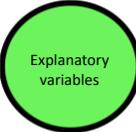 Explanatory variables | Temperature ($Temp_i$) | Unknown measurement error | Explicit estimation of the variation introduced to the explanatory variable by measurement error ($\sigma_\eta^2$) e.g. $$Temp_i^* = Temp_i + \eta_i$$ $$\eta_i \sim N(0, \sigma_\eta^2)$$ Where $Temp_i^*$ is the true temperature and $Temp_i$ is the observed temperature |
| 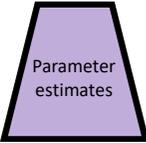 Parameter estimates | Estimates of: Intercept ($\hat\beta_0$), slope of relationship ($\hat\beta_1$), and variance of the error ($\hat\sigma^2$) | A range of values could be plausible as the 'true' parameter value for any parameter | Standard error/confidence interval |

| 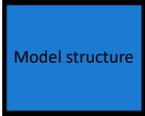 Model structure | The structure of the equation | Other processes could influence change in plant height e.g. rainfall, or it could even be a non-linear relationship with temperature | Comparison of alternative formulations e.g. non-linear structure or additional explanatory variables |

*Model type definitions*

While the term 'model' is often used as shorthand in specific disciplines to refer to the predominant model type, here we take a broader approach and identify three key types of model, which are defined below.

- *Statistical models*. A mathematical model that represents a data generation process. These models aim to say something about a population of interest based on data from a sample taken from that population. Examples of statistical models are linear regression, t-test, analysis of variance.
- *Dynamical models (can also be called mechanistic).* A mathematical model based on fundamental understanding of natural processes (e.g. physical or biochemical laws). Dynamical models take the form of a series of equations based on physical and biochemical principles that are used to understand systems where there is change, growth, or development e.g. the climate system. They can be used to fit to or explain observed data or model output. Often applied to dynamical systems.
- *Theoretical models*. Can be similar in motivation to dynamical models but contain no observed data (even in the form of model output to represent a real variable). These models are a theoretical exercise to test a particular idea. They are used to progress theory around certain processes.

Qualitative models were assessed but were excluded from this analysis as the framework is designed for quantitative models.

## S2 Development of Audit Strategy

*Initial assessment criteria:*

The first iteration of the data collection table was used to conduct an audit on 10 papers from the Journal of Animal Ecology from the September 2019 issue. The trial was conducted by all eight members of the Ecology team and all reviewers audited the same papers. The results of the trial were compared using percentage agreement (calculated as the number of reviewers with the same answer over the total number of reviewers). The consistency of results at this stage was low (often < 70% agreement). Agreement was particularly low for the section relating to an unobserved state.

The preliminary data collection table was split into two sections, one focused on an explanatory model and the second on predictions, if they occurred. Within each section there were two mini-tables, the first covered a state, or unobserved, process and the second covered the observed process. For each process questions were asked about what form each source component took and whether the associated uncertainty was reported.

| Type of response | Component | | It is used for prediction? | |
|---|---|---|---|---|
| Should there be a state? | | | Predicted state (unobserved) | **What is the state response** Response uncertainty **What are the state explanatory vars** Explanatory uncertainty **What are the state parameters** Parameter uncertainty **What is the state model** Model uncertainty |
| State (unobserved) | **What is the state response** Response uncertainty **What are the state explanatory vars** Explanatory uncertainty **What are the state parameters** Parameter uncertainty **What is the state model** Model uncertainty | | | |
| | | | Predicted observation | **What is the observed response** Response uncertainty **What are the observation explanatory vars** Explanatory uncertainty **What are the observation parameters** Parameter uncertainty **What is the observation model** Model uncertainty |
| Observation | **What is the observed response** Response uncertainty **What are the observation explanatory vars** Explanatory uncertainty **What are the observation parameters** Parameter uncertainty **What is the observation model** Model uncertainty | | | |
| | | | Type of uncertainty communication e.g. visual, numeric, text | |
| Type of uncertainty communication e.g. visual, numeric, text | | | In main text? | |
| In main text? | | | Uncertainty included in conclusion? | |
| Uncertainty included in conclusion? | | | Multiple analyses reported? | |

| Component | Average percentage agreement |
|---|---|
| Observed Response | 48 |
| Observed Explanatory variable | 73 |
| Observed Parameter estimates | 37 |
| Observed Model structure | 50 |

FIGURE 1: a) Original table, b) Agreement table from first trial

Following the trial audit, all eight reviewers attended a three-hour workshop to give feedback on the table and discuss the inconsistencies. After the workshop, a new data collection table was drafted to streamline the data collection process and ensure results could be repeatable. During the initial trial audit, we found that the preliminary table had been too complex and with too high a focus on state vs observation processes. In streamlining the data collection table to focus more directly on each source of uncertainty and report more details on how uncertainty was presented, we ensured easier and more consistent data collection and a framework that was more easily applicable to other fields.

The revised data collection table is shown in Figure 2. We reduced the number of questions down to nine numbered questions with seven answer columns. The first four questions covered the four sources of model-based uncertainty identified in our framework. Questions five and six covered whether there was an explicit model for an unobserved process, or state. Questions seven and eight covered whether predictions were made and if the associated uncertainty was reported. Question 9 assessed whether any quantified uncertainty was discussed in the discussion or conclusions. Each of the seven answer columns collected a different bit of information about the question, full details are shown in the Instructions document in Section X.

*Refined assessment criteria:*

A second trial, using the revised data collection table of questions (see Figure 2), was conducted on five papers from Ecology and Evolution from 2019 (this was a journal that was not included in the final audit to give an independent test of the method). This second trial was conducted by four of the original eight reviewers, due to time constraints of the other four.

| Number | Questions | Paper | Initials | Answer | Details | Location | Presentation | Model Type |
|---|---|---|---|---|---|---|---|---|
| 1 | Is uncertainty in the response included? | | | | | | | |
| 2 | Is uncertainty in the explanatory variables included? | | | | | | | |
| 3 | Is uncertainty in the parameter estimation included? | | | | | | | |
| 4 | Is uncertainty in the model structure included? | | | | | | | |
| 5 | Is there an explicit model for a process that was not observed? | | | | | | | |
| 6 | Is uncertainty in this unobserved element included? | | | | | | | |
| 7 | Do they predict? | | | | | | | |
| 8 | Is uncertainty in the prediction included? | | | | | | | |
| 9 | Is uncertainty discussed in the discussion? | | | | | | | |
| 1 | Is uncertainty in the response included? | | | | | | | |
| 2 | Is uncertainty in the explanatory variables included? | | | | | | | |
| 3 | Is uncertainty in the parameter estimation included? | | | | | | | |
| 4 | Is uncertainty in the model structure included? | | | | | | | |
| 5 | Is there an explicit model for a process that was not observed? | | | | | | | |
| 6 | Is uncertainty in this unobserved element included? | | | | | | | |
| 7 | Do they predict? | | | | | | | |
| 8 | Is uncertainty in the prediction included? | | | | | | | |
| 9 | Is uncertainty discussed in the discussion? | | | | | | | |

| **Question number** | **Average percentage agreement** |
|---|---|
| 1 | 80 |
| 2 | 100 |
| 3 | 75 |
| 4 | 80 |
| 5 | 92 |
| 6 | 75 |
| 7 | 100 |
| 8 | 100 |
| 9 | 92 |

FIGURE 2: a) Revised table, b) Agreement table second trial

Agreement from the second trial was much higher (from 49% across all questions in trial 1 to 88% across all questions in trial 2) but not perfect, although absolute perfection is unlikely to be possible. The main continued areas of inconsistency arose from choosing No rather than NA for some sections. In response, a checks section was added to the spreadsheet for data entry for the final set of criteria. In addition, the detailed instructions document was refined and expanded and can be found here: in Section 4.

## S3 The Systematic Audit

*Search Strategy:*

Unlike typical Systematic Audits, we did not use search strings to target research articles relevant to our research question. Instead, we took **all** research articles from the last few issues from 2019 (the exact number depended on the number of research papers per issue and how often issues were released – the aim was for 30-50 papers per journal), of a population of academic journals across a broad-range of scientific disciplines. Our list of target scientific journals was built by consultation during workshops with the full audit team and had to meet the following criteria to be included:

- English language
- Some original research papers per issue
- International audience
- Field leading
- Frequently publish papers that use statistical models

While all original research papers from each issue of these journals were to be assessed, only those that included a statistical or mathematical analysis involving a model (it did not need to include observed data) were included in the audit.

Article types excluded from these analyses were determined by the following exclusion criteria:

- book reviews
- editorials
- in-focus pieces
- narrative reviews with no analyses
- no statistical, dynamical, or mathematical model
- purely descriptive work
- meta-analyses and systematic reviews
- models on animals (Health sciences only)

44 papers were excluded from the final assessment. They are summarised in the table below.

| Paper | Field | Reason for exclusion |
|---|---|---|
| Abimbola 2019 | Health Sciences | No model |

| Reference | Field | Classification |
|---|---|---|
| BarlowM 2019 | Climate | Review |
| Bisson 2019 | Health Sciences | Editorial |
| Burson 2019 | Ecology | No model |
| Carvalho D 2019 | Climate | No model |
| Chen J 2019 | Climate | No model |
| Chen X 2019 | Climate | No model |
| Choi Y 2019 | Climate | No model |
| Choi YW 2019 | Climate | No model |
| Dasandi 2019 | Political Science | No model |
| Desai 2019 | Health Sciences | Method paper no primary research |
| DiCesare 2019 | Health Sciences | Review |
| Fernandez-Albert 2019 | Neuroscience | Challenging paper |
| Findell K 2019 | Climate | No model |
| Han Z 2019 | Climate | No model |
| Hart N 2018 | Climate | Descriptive |
| Herzog and Zacka 2019 | Political Science | No model |
| Hobert 2019 | Political Science | No model |
| Jensen 2019 | Political Science | No model |
| Kelley 2019 | Political Science | No model |
| Kenyon 2019 | Health Sciences | No model |
| Luo M 2019 | Climate | No model |
| Maharana P 2019 | Climate | No model |
| Mildenberger 2019 | Political Science | Descriptive |
| Mwangome 2019 | Health Sciences | Editorial |
| Nie Y 2019 | Climate | No model |
| Ning G 2019 | Climate | No model |
| Papanicolas 2019 | Health Sciences | No model |
| Pinheiro H 2019 | Climate | No model |
| Porto da Silveiral 2019 | Climate | No model |
| Rodriguez-Villa 2019 | Health Sciences | Review |
| Seely 2019 | Neuroscience | Review |
| Shu Q 2019 | Climate | No model |
| Si D 2019 | Climate | No model |
| Tickell 2019 | Health Sciences | Review |
| Wolton 2019 | Political Science | No model |
| Yu Y 2019 | Climate | Qualitative |
| Yuan C 2019 | Climate | No model |
| Zhang 2019 | Health Sciences | Animal study |
| Zhang S 2019 | Climate | No model |
| ZhangRH 2019 | Climate | No model |
| Zhao H 2019 | Climate | No model |
| Zhou 2019 | Evolution | No model |
| Zhou X 2019 | Climate | No model |

*The final list of journals used in the audit were:*

| Field | Journal |
|---|---|
| Ecology | Journal of Animal Ecology |
| Ecology | Ecology |
| Molecular Ecology/Evolution | Molecular Ecology |
| Molecular Ecology/Evolution | Evolution |
| Climate Science | Journal of Climate (Volume 32, Issues 23 and 24) |
| Climate Science | Climate Dynamics (Volume 53, Issues 9-11) |
| Neuroscience | Nature Neuroscience |
| Neuroscience | The Journal of Neuroscience |
| Political Science | British Journal of Political Science |

| Political Science | American Journal of Political Science |
| Health Sciences | British Medical Journal |
| Health Sciences | BMC Medicine |
| Oceanography | Journal of Geophysical Research: Oceans |
| Oceanography | Journal of Physical Oceanography |

*Extracting papers from the journals:*

There were no fixed number of issues chosen for each journal, instead, enough issues were chosen to be able to extract around 50 original research papers per journal. This led to a total of 90-110 papers per field being extracted for audit.

Once research papers were extracted, they were randomly assigned to reviewers in each field. Each individual reviewer then screened their assigned papers for presence of a model and excluded any article types listed in the 'exclusion criteria'.

In total, we audited 50-110 papers per field. The exact number of papers audited was dependent on; time constraints of reviewers (this was particularly challenging due to increased pressures from the coronavirus pandemic), desired breadth of coverage of the field, and number of team members per field.

*Reviewer training:*

The first phase of the systematic audit was a reviewer training phase. This was conducted within each field team to ensure that all reviewers were familiar and comfortable with the instructions and framework, that all reviewers were applying the criteria consistently, and that the framework was applicable to papers from each field included in this audit.

The reviewer training phase consisted of all reviewers in each field auditing the **same four-five** papers. These papers were taken from one of the target journals for each field also from 2019 (but independent of the articles to be used for the main audit). The papers for the training phase were selected by choosing the first four-five papers in the target issue.

For Evolution/Molecular Ecology and Health Sciences the training phase was run slightly differently. Instead of the four-five papers being taken from a single journal two papers were selected from each of the two journals chosen for these fields. This tweak to the protocol was applied as the journals in these fields each have a substantially different style. In addition, the journal 'Molecular Ecology' has a sub-section structure, therefore, to ensure breadth of training papers, one paper was selected from the first subsection and one from second in the September issue of this journal. Both papers were the first papers listed in each sub-section.

Once all reviewers had completed the training audits their tables of results were sent to the project leader (E. G. Simmonds). The project leader then analysed the results by calculating percentage agreement scores in the same way as described above. A meeting was then held with each field team including the project leader to discuss the results of the training, answer any reviewer questions, address the causes of any low agreement, and ensure that the instructions were clear and easy to follow. The presence of the project lead in all of these field specific meetings ensured consistent interpretation and implementation of the audit framework across fields.

Where low agreement was noted during this training phase (low agreement defined as < 70% agreement), the causes of this were discussed in the summary meeting and a consensus achieved to resolve the cause of the inconsistency, to avoid discrepancies in future, and ensure between-reviewer consistency.

Following the first set of reviewer training (for the Ecology, Oceanography, Neuroscience, and Political Science teams), the instructions were updated to be more detailed and reflect the concerns raised during these meetings. This step ensured that any lack of clarity was immediately addressed and communicated to all members of the project, further improving between-reviewer consistency. Working documents of definitions of model types and uncertainty types were also created so if new model types or uncertainty types were encountered during the audit, these could be added in a way clearly visible to all reviewers.

The Health Sciences team chose to conduct a second round of reviewer training following the first in order to be sure of improved agreement.

*Main audit phase:*

The reviewers within each field audited their assigned papers using the data collection table and updated instructions.

A protocol was in place to deal with papers that any reviewer found challenging, or if they did not feel confident that their coding was repeatable.

*Protocol for dealing with challenging papers (this was introduced in November 2020, in response to a need identified by the Ecology team and was implemented for all teams):*

If during the audit any reviewer finds a paper challenging where they are unsure of answers, the following protocol will be initiated:

- The reviewer will send the paper to the contact person for their team and it will be distributed to other members (min of one but up to three) of their field team to assess if they agree with the coding by the focal reviewer.
- The code agreement can either be assessed by independently coding the challenging paper, or assessing the original reviewer's coding.
- Both/all (if >2) reviewers of each paper will meet after the audits are completed to reach a consensus on how to code these challenging papers. Agreement between all reviewers of each paper will be assessed and if no clear consensus can be reached, all team members will discuss until an agreement is reached.
- If paper is too challenging for anyone - removed from analysis as not confident of results (n=1 in neuroscience - **Fernandez-Albert et al 2019**)

A summary meeting was also held for each team to check all members had audited consistently and still agreed on the interpretation of the audit framework. If any differences in interpretation were highlighted, the project leader decided on the correct interpretation and audits were rechecked following the discussion, updated, and resubmitted to the project leader.

All final results were sent to the project leader following these field specific summary meetings.

**Data cleaning**

After submission of final results, all data tables were inspected manually first to:

- Remove any blank lines between papers in the table
- Remove example papers
- Correct any obvious spelling mistakes
- Ensure that Paper and Initials columns were filled in for all 9 questions (sometimes only the first was used)
- Check for any Excel errors e.g. increasing year for the Paper ID (2019,2020,2021)
- See if any checks had been tagged as violated to go back to reviewer for correction

After manual inspection the data were then imported to R and further checks initiated:

- Check 1 = when Answer = No, Location is NA
- Check 2 = when Answer to Number 1-4, 6 or 8-9 is Yes that Details are not NA
- Check 3 = when Answer = No, Presentation = NA
- Check 4 = when Location is not NA, presentation is not NA
- Check 5 = if the Answer to 5 or 7 is "No" then Answer to 6 and 8 = "NA"
- Check 6 = if Answer for 1-4 is "No", then Answer for 9 = "NA"

If any of these checks are violated the individual rows are checked manually. If the error is obvious and objective e.g. a clear spelling error, this is corrected by the lead author. If there is any ambiguity in the cause of the error, the row is highlighted for the original reviewer and they are asked to check.

If on reading results, any seemed surprising e.g. no parameter uncertainty for a statistical model despite uncertainty being considered for the response and covariates, these papers were reassessed by the original reviewer in discussion with the project leader, or by the project leader alone to ensure consistency across the whole project.

Other errors and how they are dealt with:

| Error | Solution |
|---|---|
| Wrote out full word when it should be an acronym | Overwrite to the acronym in R |

| Spelling error | Overwrite in R |
|---|---|
| Specific location rather than code of Main, SI etc or synonym e.g. appendix | Corrected to the expected code - all cases were unambiguous e.g. table 1 |
| Inclusion of type of uncertainty that was not appropriate e.g. AIC for model structure | Paper was rechecked by lead author and if no qualifying uncertainty type for the component was identified this was corrected to Answer = No |
| More than one answer when it should just be one e.g. "Main/SI" | Corrected to "Main" only |
| Syntax errors e.g. capitalisation or inconsistent use of spaces | Corrected in R |

## S4 Detailed reviewer instructions

*Excluded article types:*

- book reviews
- in-focus pieces
- narrative reviews with no analyses
- purely descriptive work
- meta-analyses and systematic reviews
- no original analysis/model

*Challenging papers protocol:*

Keep a note of challenging papers, they will be dealt with following this protocol:

If during the audit any reviewer finds a paper challenging where they are unsure of answers, the following protocol will be initiated:

- The reviewer will send the paper to the project leader and it will be distributed to another member or members of their field team to audit independently
- All reviewers in the field team will meet after the audits are completed to reach a consensus on how to code these challenging papers. They will be presented by the two reviewers

Table of which model to choose per field. This is slightly different for each field with the aim to typically choose the 'main' model of a paper in a consistent and repeatable manner based on standard format of papers in each field.

A model is defined here as something that draws inference beyond the sample data (this will include ALL theoretical models but exclude any descriptive statistics) and is explicit in the methods or models part of the paper.

| Field | Which model? | Comments |
|---|---|---|
| Ecology | Last model in methodology | |
| Neuroscience | First model in methodology | |
| Politics | First model in methodology | |
| Climate Science | First model mentioned in abstract | Must also run analyses in the focal paper i.e. not only use existing model output. If the first model does not meet this criteria - keep going through abstract in order until one does. |
| Oceanography | Last model in methodology | |
| Evolution | Last model in methodology | |
| Geography | Last model in methodology | |
| Health Science | Start at 1, if this criteria is not available, move to 2 etc.<br>1. "Primary analysis"<br>2. Analysis of the "primary outcome"<br>3. First explicit model after descriptive analysis | |

*Detailed instructions for the data collection table:*

| # | Question | Explanation of the question | Answer (expected responses) | Details (expected responses) | Location of uncertainty reporting (expected responses) | Presentation of uncertainty (expected responses) | Model type |
|---|---|---|---|---|---|---|---|

| | | | | | | | | |
|---|---|---|---|---|---|---|---|---|
| **1** | Is uncertainty in the response included in the focal model? | This question asks you to focus on the response variable of the model you are assessing and report if any uncertainty in that response is included and then reported in some way (i.e. so you can actually identify it) | **Yes/No/NA**<br><br>**Yes** if there is uncertainty reported and it is accounted for in the focal model e.g. measurement error is quantified, the variable is corrected and then used in the model. A Cox model with censoring would be a Yes. If a random effect model is used where measurement error can be distinguished from other patterns, this is a Yes answer as well. (Should be explained in the paper that the random effect structure is used to look at uncertainty).<br><br>**No** if there is no uncertainty reported or if it is not explicitly accounted for in the model. This should be the answer even if the uncertainty CAN'T be quantified. That will be covered in the details column. A linear regression with normal noise and No extra steps will be a No answer here.<br><br>**NA** should be used when the component is not present in the model.<br><br>If there are multiple responses and some have uncertainty reported use **Yes** but write **incomplete** in comments. | If **ANSWER = Yes** this should cover the type of uncertainty that has been reported e.g. Standard error (**SE**), Bootstrap interval (**BI**), Credible interval or confidence interval (**CI**), measurement error (**ME**), observation error (**OE**), a state model (**State**).<br><br>Full list of uncertainty types below.<br><br>If **ANSWER = No** this can explain if the uncertainty is Missing (**MISS**), Impossible (**IM**), or no uncertainty in the variable i.e. uncertainty is not relevant for this variable (**NU**), **Not propagated** (if mentioned elsewhere but not included in model)<br><br>Can also include: **NA, I don't know, can't tell**.<br><br>IF ANSWER = NO this can explain if the uncertainty is Missing (**MISS**), Impossible (**IM**), or no uncertainty in the variable i.e. uncertainty is not relevant for this variable (**NU**), I expect NU to be rare, **Not propagated** (if mentioned elsewhere but not included in model) | **Main** = in the main article, **SI** = in supporting information or appendix, **Data** = with a data publication, **Citation** = in another paper that is cited (for citation the focal paper must indicate the uncertainty is reported at the citation and also use it in the model.<br><br>Only give one answer: If multiple apply default to the most accessible (i.e. in order presented here) | Here give details of how it was reported e.g. **visual** (in a figure), **text** (in the main text as words - this can just be authors saying "we did X to quantify uncertainty in Y" but never giving a number), **numeric** (as a number either in a table or text), other.<br><br>Combinations can be used e.g. visualnumeric or visualtext. Please stay in order of **visual, text, numeric**<br><br>Report **all** presentation techniques used.<br><br>Make sure text is selected if the authors write something like "we account for uncertainty in X with Y" just saying we ran a model is not enough for text. | Only answer in this first row for the whole model:<br><br>qualitative, dynamical, statistical, theoretical |

| | | | | Yes/No/NA | If **ANSWER = Yes** this should cover the type of uncertainty that has been reported e.g. Standard error (**SE**), Bootstrap interval (**BI**), Credible interval or confidence interval (**CI**), measurement error (**ME**), observation error (**OE**), a state model (**State**). | | |
|---|---|---|---|---|---|---|---|
| **2** | Is the uncertainty in the explanatory variables included in the focal model? | This question asks you to focus on the explanatory variables of the model you are assessing and report if any uncertainty is included and reported in some way | | **Yes** if there is uncertainty reported and it is accounted for in the focal model e.g. measurement error is quantified, the variable is corrected and then used in the model.<br><br>**No** if there is no uncertainty reported and if it is not explicitly accounted for in the model. This should be the answer even if the uncertainty CAN'T be quantified. That will be covered in the details column.<br><br>**NA** should be used when the component is not present in the model.<br><br>If there are multiple explanatory variables and some have uncertainty reported use **Yes** but write **incomplete** in comments. | Full list of uncertainty types below.<br><br>If **ANSWER = No** this can explain if the uncertainty is Missing (**MISS**), Impossible (**IM**), or no uncertainty in the variable i.e. uncertainty is not relevant for this variable (**NU**), **Not propagated** (if mentioned elsewhere but not included in model)<br><br>Can also include: **NA, I don't know, can't tell**. | **Main, SI, Data, Citation**<br>(same as response instructions) | **visual, text, numeric**<br>(same as response instructions) |
| **3** | Is the uncertainty in the parameter estimation included (reported in the paper)? | For the model you are looking at, do they report uncertainty in their estimates or the values of the unknown parameters? | | Yes/No/NA<br><br>**Yes** if there is uncertainty reported and it is accounted for in the focal model e.g. measurement error is quantified, the variable is corrected and then used in the model.<br><br>**No** if there is no uncertainty reported and if it is not explicitly accounted for in the model. This should be the answer even if the uncertainty CAN'T be quantified. That will be covered in the details column.<br><br>**NA** should be used when the component is not present in the model.<br><br>If there are multiple parameters and some have uncertainty reported use **Yes** but write **incomplete** in comments. | If **ANSWER = Yes** this should cover the type of uncertainty that has been reported e.g. Standard error (**SE**), Bootstrap interval (**BI**), Credible interval or confidence interval (**CI**), range of values used as parameter values (**range**).<br><br>Full list of uncertainty types below.<br><br>If **ANSWER = No** this can explain if the uncertainty is Missing (**MISS**), Impossible (**IM**), or no uncertainty in the variable i.e. uncertainty is not relevant for this variable (**NU**), **Not propagated** (if mentioned elsewhere but not included in model)<br><br>Can also include: **NA, I don't know, can't tell**. | | |
| 4 | Is the uncertainty in the model structure included (reported in the paper)? | For the model you are looking at, do the authors consider how other model types/structures etc might have performed? Or give any range of answers? This question goes beyond the focal model as it can include comparison to other models | **Yes/No/NA** **Yes** if there is uncertainty reported and it is accounted for in the focal model e.g. measurement error is quantified, the variable is corrected and then used in the model. **No** if there is no uncertainty reported and if it is not explicitly accounted for in the model. This should be the answer even if the uncertainty CAN'T be quantified. That will be covered in the details column. **NA** should be used when the component is not present in the model. Can be Yes for sensitivity analysis that looks at model structure but not for those looking at sub-groups only e.g. removing smokers. | If **ANSWER = Yes** this should cover the type of uncertainty that has been reported e.g. Model comparison (**MC**), model averaging (**MA**), Ensemble (**Ensemble**) Full list of uncertainty types below. If **ANSWER = No** this can explain if the uncertainty is Missing (**MISS**), Impossible (**IM**), or no uncertainty in the variable i.e. uncertainty is not relevant for this variable (**NU**), **Not propagated** (if mentioned elsewhere but not included in model) Can also include: **NA, I don't know, can't tell**. Model selection also does not count unless results compared or combined i.e. **model averaging** | | | |
| 5 | Is there an explicit model that maps an unobserved response to observed data? (response) | If the response being modelled is not the same thing the authors had data for then do they map the data and desired response to each other with a model? An example of this would be survival analysis. The data is often recapture (whether you saw an individual again or not), but the model tries to explain survival. Therefore, you have a model that includes both recapture and survival probability explicitly - this would give a **Yes** here. Might expect many **No** answers. | **Yes/No** | Give some detail e.g. the process that is unobserved like: **survival**. | NA | NA | |
| 6 | Is uncertainty in this element included? | If they model something unobserved, do they explicitly account the uncertainty from this? And the uncertainty associated. If the question above was Yes, the answer to this will be the same as Qu1. | **Yes/No/NA** If the answer to Qu5 is No, the answer here should always be NA | Details of uncertainty type e.g. **state** = full state model with reported uncertainty in the response, **ME** = measurement error, **OE** = observation error, **other, NA, I don't know, I can't tell**. | **Main, SI, Data, Citation** (same as response instructions) | **visual, text, numeric** (same as response instructions) | |
| 7 | Do they predict? | Are any predictions made during the final analysis or from the | **Yes/No** | | NA | NA | |

| | | | | | | |
|---|---|---|---|---|---|---|
| | | model in the final analysis? Predictions are defined as estimating/calculating values of the response outside of the data used in the model. Temporal element of past or future is not important here. Projections would count **if** generated by the focal model in the focal paper. | | | | |
| 8 | Is uncertainty in the prediction included? | Only answer if question above = Yes | **Yes/No/NA** If the answer to Qu7 is No, the answer here should always be NA | Type of uncertainty e.g. prediction interval (**PI**), **SE**, **CI**, **Range** = range of scenarios, If **ANSWER = No** can explain why e.g. **not propagated** Also, **NA, I don't know, I can't tell** | | |
| 9 | Is uncertainty from the f**ocal model** discussed in the discussion? | Is any uncertainty discussed in the discussion? This focuses on the discursive part of a paper, often the 'discussion or conclusion' section. You should look for whether the meaning of the uncertainty is included or any reference to how it impacts conclusions. This is not the same as the raw reporting of uncertainty in a results section. **This question concerns any uncertainty that was quantified above, not unquantified uncertainty.** Potential search words to help identify uncertainty in the discussion: uncertain, overlap, not clear, clear, range, ± | **Yes =** uncertainty is discussed in the paper beyond just reporting. i.e discuss implications. **No** = the uncertainty is not discussed after results presented. **NA** = to be used if no uncertainty was quantified. | Which part e.g. **Response, explanatory, parameter, model, unobserved, prediction, all, none**. | **Main, SI, Data, Citation** (same as response instructions) | **visual, text, numeric** (same as response instructions) |

*Uncertainty type definitions:*

This whole audit focuses on quantitative uncertainty.

- **Standard error** (**SE**): a standard error from a statistical model. This can be a standard error of a mean, of a difference, etc.
- **Bootstrap interval (BI)**: an interval of confidence (say 95% but could be other levels) obtained by bootstrapping (resampling data and re-running an analysis).
- **Credible interval or confidence interval (CI)**: an interval indicating plausible values usually of an unknown parameter but can also cover values of a latent response or explanatory variable.
- **Measurement error (ME)**: error or bias coming from the measurement of a variable. Should be quantified using an error in the variables model or other appropriate model.
- **Observation error (OE)**: error from observing a process. Should be quantified using an observation model.
- **A state model (State)**: coupled with an observation model, this part models the underlying process of interest and is linked to the observation model.
- **Multiple scenarios (scenarios)**: running an analysis to cover multiple scenarios to give a range of results indicating uncertainty in either model structure, parameters, or variables.
- **Model averaging (MA):** taking a selection of models and averaging parameter estimates. This represents uncertainty introduced by model structure.
- **Model ensemble (En)**: taking a collection of models with different structures and presenting results from all of them simultaneously with uncertainty bands reflective of all models in the ensemble. Represents some uncertainty in model structure.
- **Model comparison (MC):** running models with different structures or parameter values or inputs and comparing the results to give an indication of the uncertainty introduced by the varied component.
- **Prediction interval (PI)**: an interval indicating a level of uncertainty in a prediction (usually of a response).
- **Range:** model run using a variety of values for that source e.g. using a sequence of parameter values or selecting parameter values randomly from distributions.
- **Standard deviation (SD):** but only of model outputs or estimated parameters, not of observed data (in this later case it just shows variability).
- **Censoring (Censoring):** for example for survival data where the outcome of certain individuals is not known.
- **Full distribution:** using a full posterior distribution from a Bayesian analysis.
- **Hierarchical partitioning:** partitions variance explained among independent variables through assessing the contributions of all orders of the variables and calculating a partition of the variance. Calculates the contribution of each independent variable separately and in conjunction with other variables.

**Data availability:** All data and code used in the analysis will be made available on Zenodo repository with a dedicated DOI.